\documentclass[aps,pra,reprint,superscriptaddress]{revtex4-1}
\usepackage{hyperref,graphicx,color}

\newcommand{\QUOTE}[3]{\begin{itemize} \item[] \emph{``#1"} 
\hfill (#3) \end{itemize}}

\newcommand{\figFRAMEWORK}{
\begin{figure}[t]
\includegraphics[width=\columnwidth]{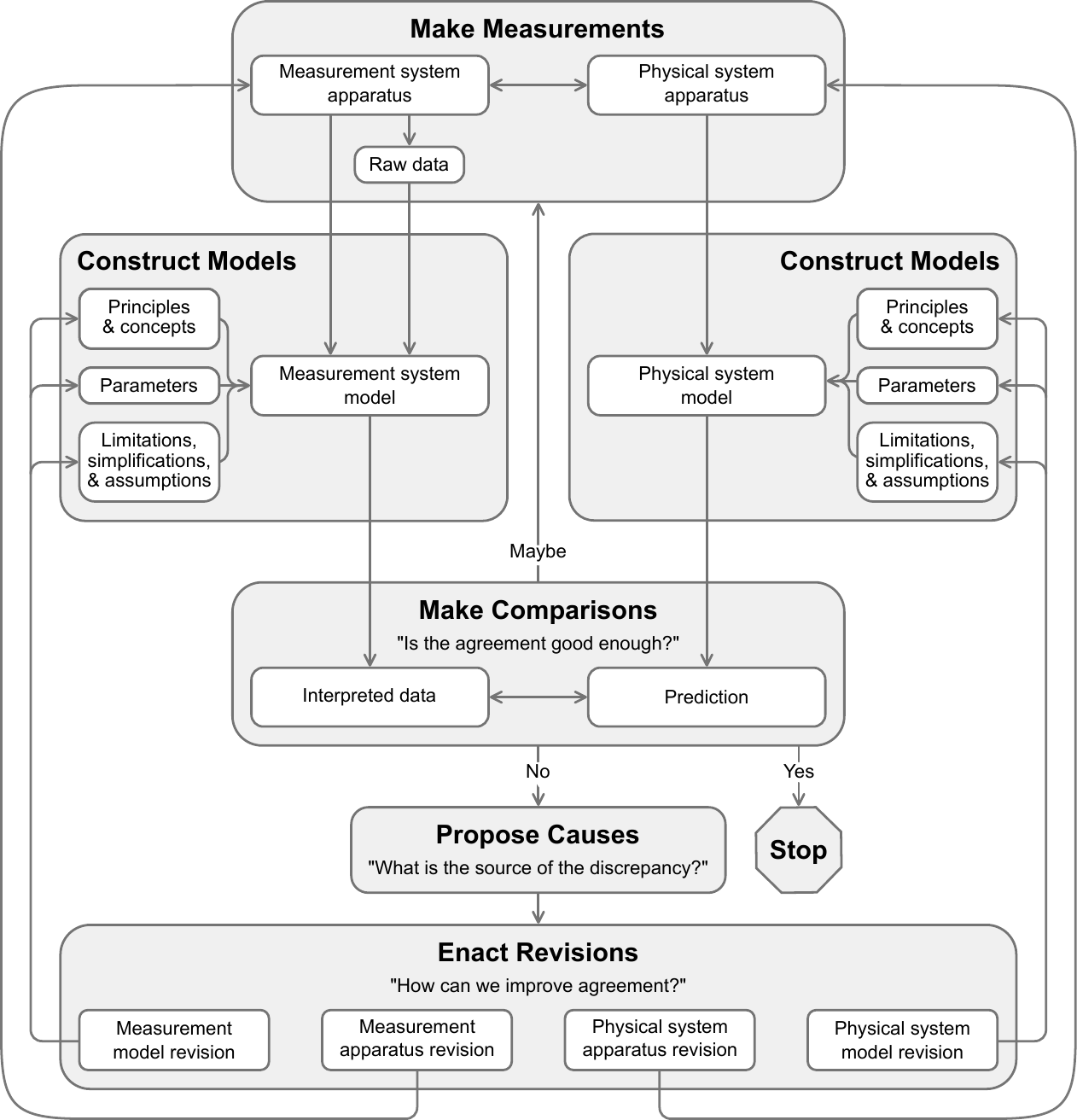}
\caption{\label{fig:framework}Modeling Framework. The diagram here differs slightly from that of Ref.~\cite{Zwickl2015}. For example, this version includes a path from \emph{make comparisons} to \emph{make measurements}.}
\end{figure}
}

\begin{document}

\title{Instructor perspectives on iteration during {upper-division} optics lab activities}

\author{Dimitri R. Dounas-Frazer}
\affiliation{Department of Physics, University of Colorado Boulder, Boulder, CO 80309, USA}

\author{Jacob T. Stanley}
\affiliation{Department of Physics, University of Colorado Boulder, Boulder, CO 80309, USA}

\author{H. J. Lewandowski}
\affiliation{Department of Physics, University of Colorado Boulder, Boulder, CO 80309, USA}
\affiliation{JILA, National Institute of Standards and Technology and University of Colorado Boulder, Boulder, CO 80309, USA}

\date{\today}

\begin{abstract}
{Although developing proficiency with modeling is a nationally endorsed learning outcome for upper-division undergraduate physics lab courses, no corresponding research-based assessments exist. Our longterm goal is to develop assessments of students' modeling ability that are relevant across multiple upper-division lab contexts.} To this end, we interviewed 19 instructors from 16 institutions about optics lab activities that incorporate photodiodes. Interviews focused on how those activities were designed to engage students in some aspects of modeling. We find that, according to many {interviewees}, iteration is an important aspect of modeling. In addition, interviewees described four distinct types of iteration: revising apparatuses, revising models, revising data-taking procedures, and repeating data collection using existing apparatuses and procedures. We provide examples of each type of iteration, and discuss implications for the development of future modeling assessments.
\end{abstract}

\maketitle


\section{Introduction}

The ability to model physical systems is a nationally recognized learning outcome for undergraduate physics lab courses~\cite{AAPT2015}. At the introductory level, multiple approaches have been designed to engage students in the process of developing and revising models (e.g., Refs.~\cite{Brewe2008,Etkina2007,Gandhi2016}). At the upper-division level, {we have} participated in course transformation efforts that emphasize model-based reasoning in both advanced lab~\cite{Zwickl2013,*Zwickl2014} and electronics lab~\cite{Lewandowski2015,Stanley2017arXiv} courses at the University of Colorado Boulder. These course transformations coincided with the development of the Modeling Framework for Experimental Physics (hereafter, ``the Modeling Framework"), which describes the cyclic process that physicists employ when refining models and apparatuses~\cite{Zwickl2015}. Recently, the Modeling Framework has been used to characterize students' approaches to completing both optics~\cite{Zwickl2015} and electronics~\cite{Dounas-Frazer2016a} tasks. Here, we use it in a new capacity: to determine whether and how instructors prioritize the iterative nature of modeling in upper-division lab courses.

Though modeling is an important learning outcome for upper-division labs, no standardized measures of students' modeling abilities exist in this context. Accordingly, we are in the process of developing assessments of students' model-based reasoning during experimental physics tasks. {Similar to} the development of conceptual assessments~\cite{Engelhardt2009,Wilcox2015}, knowing instructors' priorities for student learning is also part of establishing relevant test objectives for skill-based assessments. We are aware of two studies that focus on lab instructors' teaching goals and practices~\cite{Coppens2016a,Dounas-Frazer2017}; neither focuses on modeling. Therefore, we performed a qualitative study to explore the potential alignment of the Modeling Framework with lab instructors' goals for, and descriptions of, activities in optics labs and related courses. {In the present analysis, we focus on instructors' perspectives on the iterative aspects of modeling that arise during optics activities}.

We interviewed 19 instructors from 16 institutions across the United States about the details of optics activities that incorporate photodiodes. Collectively, interviewees described four types of iteration on these activities: (i) revising apparatuses, (ii) revising models, (iii) revising data-taking procedures, and (iv) repeating data collection with existing apparatuses and procedures in order to improve the statistical precision of a measurement. In this article, we elaborate on each type of iteration, discuss implications for instruction, and argue that assessments of lab skills should focus on experimental processes.


\section{Modeling Framework}

In the Modeling Framework, \emph{models} and \emph{modeling} are defined as follows~\cite{Zwickl2015}. Models are equations, drawings, words, or other abstract representations of real-world systems and phenomena. They are rooted in theoretical principles and concepts, and they contain simplifying assumptions that result in tractable representations with practical, but limited, explanatory and predictive power. Modeling is the dynamic process through which models and systems are refined in order to align predictions with data. The Modeling Framework was developed to describe precisely this dynamic process.

A diagram of the Modeling Framework is provided in Fig.~\ref{fig:framework}. The gray boxes correspond to modeling subtasks: \emph{make measurements}, \emph{construct models} of both the measurement and physical system apparatuses, \emph{make comparisons} between data and predictions, \emph{propose causes} for discrepancies between data and predictions, and \emph{enact revisions} to models or apparatuses of either the measurement or physical systems. The framework distinguishes between the measurement system and physical system being studied. Whereas the measurement system model is used to interpret data (e.g., a change in photodiode voltage indicates a change in light intensity), the physical system model is used to make predictions (e.g., a sample will fluoresce under some conditions, but not others). 

The Modeling Framework captures the nonlinear and recursive nature of experimentation. For example, when making comparisons, one must determine whether or not data and predictions are in ``good enough" agreement for their particular context. Depending on the level of agreement, one may stop the experiment, perform additional measurements to collect more data, or propose causes for the discrepancy. In this sense, the ``Maybe" and ``No" pathways in Fig.~\ref{fig:framework} roughly correspond to efforts to reduce statistical and systematic uncertainties, respectively. In this work, we explore how these and other iterative aspects of modeling can arise during optics lab activities.

\figFRAMEWORK


\section{Methods}

We conducted semistructured interviews with instructors of optics labs, intermediate labs, and advanced labs. Interviews were designed to probe instructors' perspectives on the role of modeling in optics activities. Our interview protocol consisted of 26 questions: 11 about departmental and course context, 13 about the details of an optics activity, 1 about the relevance of the Modeling Framework, and 1 about participants' race and gender.

Each interview focused on an optics activity of the interviewee's choice. To ensure that the activities discussed in different interviews had at least one component in common, interviewees were asked to choose an activity that incorporated one or more photodiodes. Activities spanned a variety of phenomena, such as Fresnel reflection, single-slit diffraction, and properties of photodiodes. After asking for a general overview of the activity, the interviewer described the Modeling Framework and gave the interviewee a visual representation of the framework. All subsequent questions about the activity referred to specific aspects of the Modeling Framework (e.g., ``When working on this activity, do students propose explanations for why their data and predictions don't agree? If so, can you tell me more about this?").

To solicit participation in the study, we created a database of undergraduate physics programs. Three categories of program were included in the database: large programs; all programs at Women's Colleges, Historically Black Colleges and Universities, and Hispanic Serving Institutions; and programs chosen randomly from the American Institute of Physics (AIP) roster of physics departments~\cite{AIP2016}. The database had 154 entries, about 50 from each category. Each entry in the database included information about the university and department per the Carnegie classification system~\cite{Carnegie2015} and the AIP roster, respectively. In addition, we added contact information for department chairs and relevant lab instructors; this information was publicly available on department websites. 

We solicited participation by email. We contacted everyone in the database for whom we had contact information: 150 department chairs and 64 optics instructors. In total, 19 instructors participated in our study. Based on participants' self-reported race and gender, one participant was Black and African American, one was Indian, one was Asian, and one was Caucasian with some Asian background; the other 15 participants were white or Caucasian. One participant was a woman and 18 were men.

Study participants represented 16 distinct physics departments: 10 at public universities, and 6 at private not-for-profit ones; 3 at Hispanic Serving Institutions, and 12 at Predominantly White Institutions. In nine departments, a bachelor's degree is the highest physics degree offered; two offer master's degrees, and five offer doctoral degrees. In terms of size, three departments award up to 5 bachelor's degrees per year, nine award 10 to 30 per year, and four award 60 to 100 per year. 

Interviews were conducted via videoconference. Each interview lasted 40 to 60 minutes. Audio data for each interview were recorded and transcribed. The transcripts are the data that we analyzed. The six subtasks of the Modeling Framework in Fig.~\ref{fig:framework} served as first-pass codes for a dual-pass coding process. First, the first author read through each transcript and {identified transcript excerpts} related to each subtask. Second, the same author {coded} emergent themes for each subtask. The research team collaboratively verified the appropriateness of codes assigned during both passes. Here, we report on themes related to iteration; these themes emerged when interviewees were discussing revision and comparison.


\section{Results and interpretation}

First, we describe instructors' perceptions of the importance of iteration. Then, we present descriptions of four types of iteration: (i) revising apparatuses, (ii) revising models, (iii) revising data-taking procedures, and (iv) repeating data collection. {When presenting excerpts, we identify the corresponding activity in order to convey the breadth of experiments discussed during interviews.}

When asked which aspects of the Modeling Framework were important for students to learn, {8 instructors} identified the iterative nature of modeling as important. Consider Sage and Mustard, who described {optical pumping} and {Fresnel reflection} experiments, respectively. Sage said that one major learning outcome of his course is students' view of experimentation as an iterative process, and Mustard said the recursiveness of the framework is a realistic representation of experimentation:
\QUOTE{The other thing I really liked about [the framework] is the idea of iteration. One of the biggest changes that I see---the positive changes that I see---in the students, is that they go from having a very static, fixed view of everything, that like, ``Oh, this should all be working because I'm taking a class, and it will always work." But then realizing that they need to be constantly checking and revising their understanding of the experiment and the model that they developed for how things work.}{O18}{Sage}
\QUOTE{I like that, just visually, there's one tiny fraction [of the framework] that's maybe 5\% of the space which is `Stop.' That's probably an overestimate of how things work in the lab. I wish as much as 5\% of the time we had success in the lab! I like that. I like the imagery of continually recycling into things. I think that's absolutely appropriate.}{O13}{Mustard}

Four instructors said that engaging students in iteration was not a goal of their activities. Three of these instructors articulated a tension between the number of activities covered in the course and the depth with which students could engage in any of them. For example:
\QUOTE{The way the curriculum was designed did not promote [iteration] because we're talking about an idea maybe on Tuesday, then you do a lab on Thursday, and then we talk a little bit more the next Tuesday, and then we're on to the next topic.}{O20}{Oregano}

Almost all instructors said that their activities provided students with opportunities to iteratively improve an experiment. In total, 17 people described at least one type of iteration, and 11 described multiple types. Apparatus revisions were described by 15 people. The other types of iteration were each described by 6 or 7 people.

\textbf{Apparatus revisions} included realigning optics, adding or removing optical components, revising the photodiode circuit, blocking ambient light, and changing settings on equipment. Sometimes, apparatus revisions were made in the context of troubleshooting problems. For example, Anise said that alignment was one of the most difficult aspects of a {single-photon interference} experiment, and Cardamom framed troubleshooting as a desirable aspect of a {Fraunhofer diffraction} experiment:
\QUOTE{They'll take data and, because the alignment wasn't quite right, because they hadn't zeroed things properly, etcetera, they end up getting data that they end up doing an analysis on that's not so pretty. \ldots\ The alignment is probably the trickiest thing. It should be straight forward, but it never really is.}{O17}{Anise}
\QUOTE{I think I like the troubleshooting aspect of it, also. They put it together, and it should've worked, and it didn't. Why? When you set up an experiment and it doesn't work, it almost always happens that there's more than one thing that's wrong. It's not just one thing. There's several things conspiring to confuse you, and how do you navigate your way through that?}{O03}{Cardamom}

\textbf{Model revisions} included idiosyncratic changes to models of phenomena that were specific to a particular experiment. Such revisions also involved fixing computational mistakes or accounting for nonideal aspects of photodiodes (e.g., finite active area, nonzero response time, or nonlinear voltage responses at very low or high light intensities). For example, Saffron said that some students correct erroneous equations when analyzing data on a {Fresnel reflection} experiment, and Turmeric said students calibrate their photodiode output to account for saturation effects during a {Malus's Law} experiment:
\QUOTE{Sometimes it'll be, `We didn't get our model right. We know what the answers are going to be, but when I went to plot it [on the computer], I didn't do my formula correctly. When I threw in my data that I measured, and tried to get my output points, it didn't look right.' That's an easy fix. They'll go back and fix that.}{O04}{Saffron}
\QUOTE{They don't realize that when the top is kind of flattened, it's 'cause of saturation and things like that. And then they're getting terrible results with their comparison to theory. \ldots\ And then I give them a set of neutral density filters and tell them to create a calibration curve to their photodiodes, and then apply that calibration curve to the data they continue to write.}{O09}{Turmeric}

\textbf{Revisions to data-taking procedures} involved students repeating an experiment after making changes to the way measurement equipment is used. Consider, for example, the following excerpts from interviews with Anise and Wormood (the latter of whom was describing a {Fraunhofer diffraction} experiment):
\QUOTE{They are definitely trying to work out how the equipment works, taking data, realizing there are things they didn't understand, and revising those. In a way it's probably more about revision of the procedure rather than the equipment itself, but also use of the equipment.}{O17}{Anise}
\QUOTE{Sometimes they will actually realize that they had a poor setting on some optical component in the experiment. They'll go back in, realize there's no way to recover without taking data again, and they'll just go back and take the data.}{O19}{Wormwood}

\textbf{Repeated data collection} was often described as a byproduct of revising the apparatus or procedure. However, some instructors described instances where students collected additional data using the same apparatus and procedure in order to improve the quality of a statistical analyses. For example, Cilantro described this process  in the context of a {Gaussian laser pulse-shape} activity:
\QUOTE{A lot of them do end up coming back because the first time they go through, they realize they don't take enough data points.}{O14}{Cilantro}


\section{Summary and Discussion}

Although some instructors said that engaging students in iteration was not a goal of their activities, about half said that it is important for students to learn about the iterative nature of modeling and experimentation. Moreover, almost all instructors described student engagement in at least one of four types of iteration:  (i) revising apparatuses, (ii) revising models, (iii) revising data-taking procedures, and (iv) repeating data collection. Therefore, our results suggest that iteration is a common feature of many lab activities, though it may not always be connected to an explicit learning goal.

When interpreting these findings, two limitations must be kept in mind. First, lab instructors who teach courses that do not use photodiodes were excluded from this study. Second, study participants are likely to be instructors who enjoy reflecting on their teaching practices. As a result, the results presented here do not necessarily provide a comprehensive summary of instructor perspectives on iteration across all lab course contexts.

Despite its limitations, this study has implications for teaching model-based reasoning in labs. For example, in courses whose goals include engaging students in the cyclic aspects of modeling, different activities (or different phases of the same multiweek activity) could target different types of iteration. Some activities could even blend multiple types of iteration, similar to the Malus's Law experiment described by Turmeric. In that activity, revising the model of the photodiode to account for saturation was coupled with revising the apparatus by adding neutral density filters to the optical setup. The Modeling Framework could be a useful tool for designing activities that deliberately engage students in various types of iteration; indeed, it has already been used to this end in two lab transformation efforts~\cite{Zwickl2014,Lewandowski2015,Stanley2017arXiv}.

Our findings also have implications for developing future assessments of modeling abilities in optics contexts. Because iteration is common, measuring students' competence with iterative aspects of modeling should be an objective of such assessments. To do so, test items must focus on experimental processes as well as outcomes; a similar recommendation was made in the context of assessing the ability to troubleshoot electric circuits~\cite{Dounas-Frazer2017}. Compared to multiple-choice items, coupled multiple-response items, like those used in the Colorado Upper-division Electrostatics Diagnostic (CUE)~\cite{Wilcox2015}, may be more appropriate for assessing students' reasoning when prioritizing one type of iteration over another.

Future work will explore instructor perspectives on all aspects of the Modeling Framework.

\acknowledgments This material is based upon work supported by the NSF under Grant Nos. DUE-1611868 and PHY-1125844.

\bibliography{modeling_database_170519_short}

\end{document}